\documentclass[aps,prc,twocolumn,groupedaddress]{revtex4-1}
\usepackage{amsmath}
\usepackage{bm}
\usepackage{graphicx}
\usepackage{wasysym}
\usepackage{amssymb}

\begin{document}
	
\title{Finite amplitude method on the deformed relativistic Hartree-Bogoliubov theory in continuum: The isoscalar giant monopole resonance in exotic nuclei}
\author{Xuwei Sun}
\affiliation{State Key Laboratory of Nuclear Physics and Technology, School of Physics, Peking University, Beijing 100871, China}
\author{Jie Meng}
\email{mengj@pku.edu.cn}
\affiliation{State Key Laboratory of Nuclear Physics and Technology, School of Physics, Peking University, Beijing 100871, China}
\affiliation{Yukawa Institute for Theoretical Physics, Kyoto University, Kyoto 606-8502, Japan}

\begin{abstract}
	Finite amplitude method based on the deformed relativistic Hartree-Bogoliubov theory in continuum (DRHBc-FAM) is developed and applied to study isoscalar giant monopole resonance in exotic nuclei.
	Validation of the numerical implementation is examined for $^{208}\textrm{Pb}$.
	The isoscalar giant monopole resonances for even-even calcium isotopes from $^{40}\textrm{Ca}$ to the last bound neutron-rich nucleus $^{80}\textrm{Ca}$ are calculated, and a good agreement with the available experimental centroid energies is obtained for $^{40-48}\textrm{Ca}$. For the exotic calcium isotopes, e.g., $^{68}\textrm{Ca}$ and $^{80}\textrm{Ca}$, the DRHBc-FAM calculated results are closer to the energy weighted sum rule than the calculations on the harmonic oscillator basis, which highlights the advantages of DRHBc-FAM in describing giant resonances for exotic nuclei.
	In order to explore the soft monopole mode in the exotic nuclei,  the giant monopole resonance for the deformed exotic nucleus $^{200}\textrm{Nd}$ is investigated, where the prolate shape and the oblate shape coexist.
	A soft monopole mode near 6.0 MeV is found in the prolate case, and another one near 4.5 MeV is found in the oblate case. The transition density of the soft monopole mode shows in phase or out-of-phase vibrations near the surface region,
	which is generated by quadrupole vibrations.
\end{abstract}

\maketitle

\section{Introduction}
The new generation of radioactive ion beam facilities developed worldwide have provided more and more nuclei far from the stability valley and extended our knowledge of nuclear physics from stable nuclei to exotic ones.
The exotic nuclei, in particular those near the drip-line, are loosely bound with very extended spatial density distributions.
The coupling between the bound state and the continuum by pairing correlations and the possible deformation make it difficult to describe exotic nuclei properly.

The relativistic continuum Hartree-Bogoliubov (RCHB) theory \cite{Meng_PRL_1996,Meng_PPNP_2006},
which takes into account pairing and continuum effects in a self-consistent way,
has proven to be successful in describing the ground state properties in exotic nuclei.
The RCHB theory has achieved success in reproducing and interpreting the the neutron halo in $^{11}\textrm{Li}$ \cite{Meng_PRL_1996}, predicting the giant halo in zirconium isotopes \cite{Meng_PRL_1998},
extending the boundary of nuclear chart \cite{Xia_ADNDT_2018}, etc.
To provide a proper description of deformed exotic nuclei,
the deformed relativistic Hartree-Bogoliubov theory in continuum (DRHBc) was developed
\cite{Zhou_PRC_2010,Li_PRC_2012,Zhang_PRC_2020},
with the deformed relativistic Hartree-Bogoliubov equations solved in a Dirac Woods-Saxon basis \cite{Zhou_PRC_2003}.
The inclusion of deformation facilitates the applications of DRHBc, for example,
in the predicting of the shape decoupling between the core and the halo in $^{44}\textrm{Mg}$ \cite{Zhou_PRC_2010},
and in the seeking for possible bound nuclei beyond the drip line \cite{Zhang_PRC_2021,Pan_PRC_2021}.

In order to investigate the excitations of exotic nuclei,
many-body approaches beyond the mean-field approximation should be adopted \cite{Ring_Schuck}.
For the widely used random phase approximation (RPA) method
\cite{Sun_PRC_2018,Sun_PRC_2019,Sun_CPC_2018},
calculating and diagonalizing the RPA matrix are extremely time-consuming for deformed exotic nuclei.
Instead, the finite amplitude method (FAM) \cite{Nakatsukasa_PRC_2007} is equivalent to RPA but numerically feasible.
FAM avoids the calculation of the matrix elements of two-body residual interactions and has been implemented on
Skyrme density functionals
\cite{Inakura_PRC_2009,Hinohara_PRC_2013,Kortelainen_PRC_2015}
and relativistic density functionals
\cite{Liang_PRC_2013,Niksic_PRC_2013,Sun_PRC_2017,Bjelcic_CPC_2020}.
The applications of FAM include the study of giant monopole resonance \cite{Sun_PRC_2019_2},
exotic excitation mode like pygmy dipole resonance \cite{Inakura_PRC_2011} and soft monopole mode \cite{Sun_PRC_2021},
$\beta$ decay half-lives \cite{Mustonen_PRC_2014}, and collective inertia in spontaneous fission \cite{Washiyama_PRC_2021}, etc.

As one of the fundamental excitations in a nucleus, giant resonances are small-amplitude collective vibration modes \cite{Ring_Schuck,Book_Harajeh}.
In particular, because of its close correlation with the nuclear incompressibility,
the isoscalar giant monopole resonance (ISGMR), i.e., the breathing mode of a nucleus,
has been one of the most intriguing topics in nuclear physics and astrophysics \cite{Blaizot_PR_1980}.
The nuclear incompressibility is a key parameter in nuclear equation of state (EoS),
which has important impacts on the heavy ion collision dynamics \cite{Stock_PRL_1982}
as well as astrophysical events like supernova explosions \cite{Yasin_PRC_2020}.

For exotic nuclei with a large neutron excess, a soft monopole mode may emerge, which brings new insights into the nuclear incompressibility and has become the goals for both experimental and theoretical investigations. For instance, it has been observed experimentally in $^{11}\textrm{Li}$ \cite{Fayans_PLB_1992} and $^{68}\textrm{Ni}$ \cite{Vandebrouck_PRL_2014}, and is predicted in the neutron-rich magnesium \cite{Pei_PRC_2014}, calcium \cite{Afannsjev_PRC_2015}, nickel \cite{Sun_PRC_2021}, tin \cite{Khan_PRC_2013}, and lead \cite{Khan_PRC_2013} isotopes. However, for heavy and deformed exotic nuclei, the structure and mechanism of the soft monopole mode are not clear.

Coupling to continuum is important to nuclear giant resonances, which has been shown in previous continuum RPA calculations with relativistic density functional \cite{Daoutidis_PRC_2009} and Skyrme density functional \cite{Hamamoto_PRC_2014}, as well as the continuum quasiparticle RPA calculations \cite{Nakatsukasa_RMP_2016,Matsuo_NPA_2001}.
The DRHBc theory roots in the relativistic density functional which has attracted wide attention for many attractive advantages \cite{Ginocchio_PR_2005,Liang_PR_2015},
and describes a variety of nuclear phenomena in nuclear physics successfully \cite{Meng_book_2016,Ring_PPNP_1996,Vretenar_PR_2005,Niksic_PPNP_2011,Meng_FP_2013,Meng_JPG_2015,Shen_PPNP_2019}.
Combining the advantages of DRHBc in describing exotic nuclei and the feasibility of FAM will provide
a powerful tool to investigate the impacts of deformation, pairing, and continuum effects on the giant resonances in exotic nuclei.
This paper is devoted to implementing the finite amplitude method on the deformed relativistic Hartree-Bogoliubov
theory in continuum (DRHBc-FAM) and study the isoscalar giant monopole resonance in exotic nuclei,
with special attention paid on the soft monopole mode for deformed exotic nuclei.
The paper is organized as follows.
Sec. II briefly presents the formalism for DRHBc and FAM.
The numerical details will be given in Section III.
In Sec. IV, the ISGMRs for even-even calcium isotopes will be calculated and the continuum effects will be highlighted.
In Sec. V, DRHBc-FAM will be applied to the deformed loosely bound nucleus $^{200}\textrm{Nd}$, focusing on the soft monopole mode.
Conclusions and remarks will be given in Section VI.

\section{Formalism}

\subsection{Deformed relativistic Hartree-Bogoliubov theory in continuum}
In relativistic density functional theory (RDFT) \cite{Meng_book_2016},
the energy of a nucleus at the state $|\Phi\rangle$, which is the expectation value of Hamiltonian,
can be expressed as a functional of the density $\hat{\rho}$,
\begin{equation}
\epsilon[\hat{\rho}]=\langle\Phi|\int d^3r\mathcal{H}|\Phi\rangle.
\end{equation}
For point-coupling type RDFT, the Hamiltonian density $\mathcal{H}$ is obtained from the Lagrangian density $\mathcal{L}$ \cite{Nikolaus_PRC_1992},
\begin{equation}
	\begin{aligned}
		\mathcal{L}=&
		\bar{\psi}(i\gamma_{\mu}\partial^{\mu}-M)\psi
		-\frac{1}{4}F^{\mu\nu}F_{\mu\nu}
		-e\bar{\psi}\gamma^{\mu}\frac{1-\tau_3}{2}A_{\mu}\psi\\
		-&\frac{1}{2}\alpha_s(\bar{\psi}\psi)(\bar{\psi}\psi)
		-\frac{1}{2}\alpha_V(\bar{\psi}\gamma_{\mu}\psi)(\bar{\psi}\gamma^{\mu}\psi)\\
		-&\frac{1}{2}\alpha_{TV}(\bar{\psi}\vec{\tau}\gamma_{\mu}\psi)(\bar{\psi}\vec{\tau}\gamma^{\mu}\psi)\\
		-&\frac{1}{3}\beta_s(\bar{\psi}\psi)^3
		-\frac{1}{4}\gamma_s(\bar{\psi}\psi)^4
		-\frac{1}{4}\gamma_V[(\bar{\psi}\gamma_{\mu}\psi)(\bar{\psi}\gamma^{\mu}\psi)]^2\\
		-&\frac{1}{2}\delta_V\partial_{\nu}(\bar{\psi}\gamma_{\mu}\psi)\partial^{\nu}(\bar{\psi}\gamma^{\mu}\psi)\\
		-&\frac{1}{2}\delta_{TV}\partial_{\nu}
                       (\bar{\psi}\vec{\tau}\gamma_{\mu}\psi)\partial^{\nu}(\bar{\psi}\vec{\tau}\gamma^{\mu}\psi),
	\end{aligned}
\end{equation}
in which $\psi$ and $A_{\mu}$ respectively represent the nucleon field and photon field,
the coupling constants $\{\alpha_s,\alpha_V,\alpha_{TV},\beta_s,\gamma_s,\gamma_V,\delta_V,\delta_{TV}\}$ are determined by the masses and radii of selected finite nuclei.

The single-particle Hamiltonian is the derivative of the energy functional respect to the density,
\begin{equation}
\hat{h}  =\frac{\delta\epsilon}{\delta\rho}
=\bm{\alpha}\cdot(\bm{p}-\bm{V}) + \beta (m+S) + V,
\end{equation}
which contains a scalar potential $S=\Sigma_s$,
and a vector potential $V^{\mu}\equiv(V,\bm{V})=\Sigma^{\mu}+\tau_3\cdot\Sigma^{\mu}_{TV}$,
\begin{equation}\label{vspot}
	\begin{aligned}
		&\Sigma_s=\alpha_s\rho_s+\beta_s\rho_s^2+\gamma_s\rho_s^3+\delta_s\Delta\rho_s,\\
		&\Sigma^{\mu}=\alpha_V j^{\mu}_V + \gamma_V(j^{\mu}_V)^3
		+\delta_V\Delta j^{\mu}_V
		+\frac{1-\tau_3}{2}A^{\mu},\\
		&\Sigma^{\mu}_{TV}=\alpha_{TV} j^{\mu}_{TV}
		+\delta_{TV}\Delta j^{\mu}_{TV}.
	\end{aligned}
\end{equation}

The pairing interaction is a zero-range pairing force,
\begin{equation}
V^{\textrm{pp}}(\bm{r}_1,\bm{r}_2)=
V_0\frac{1}{2}(1-P^{\sigma})\delta(\bm{r}_1,\bm{r}_2)
\Big(1-\frac{\rho(\bm{r}_1)}{\rho_{\textrm{sat}}}\Big),
\end{equation}
which leads to the pairing potential,
\begin{equation}\label{gapequ}
\Delta(\bm{r})=V_0\Big(
1-\frac{\rho(\bm{r})}{\rho_{\textrm{sat}}}
\Big)\kappa(\bm{r}),
\end{equation}
with the pairing tensor $\kappa(\bm{r})$ given in the following.

The details of the DRHBc theory with meson-exchange and point-coupling density functionals can be found in Refs.
\cite{Zhou_PRC_2010,Li_PRC_2012,Zhang_PRC_2020}.
In the DRHBc theory, the relativistic Hartree-Bogoliubov (RHB) equation reads,
\begin{equation}\label{eqrhb}
\bigg(\!\!
\begin{array}{cc}
	h-\lambda & \Delta\\
	-\Delta^* & -h^*+\lambda
\end{array}
\!\!\bigg)
\bigg(\!\!
\begin{array}{c}
	U_k\\
	V_k\\
\end{array}
\!\!\bigg)
=E_k\bigg(\!\!
\begin{array}{c}
	U_k\\
	V_k\\
\end{array}
\!\!\bigg),
\end{equation}
with the quasiparticle energy $E_k$ and corresponding spinors $U_k$ and $V_k$ as well as the chemical potential $\lambda$ taken care of the particle number conservation.
The density, current, and pairing tensor used in Eqs. \eqref{vspot} and \eqref{gapequ} can be calculated as,
\begin{equation}
\begin{aligned}
	\rho_s(\bm{r}) &= \sum_{k}V_k^{\dagger}(\bm{r})\gamma_0 V_{k}(\bm{r}),\\
	j_V^{\mu}(\bm{r}) &= \sum_{k}V_k^{\dagger}(\bm{r})\gamma_0\gamma^{\mu} V_{k}(\bm{r}),\\
	j_{TV}^{\mu}(\bm{r}) &= \sum_{k}V_k^{\dagger}(\bm{r})\tau_3\gamma_0\gamma^{\mu} V_{k}(\bm{r}),\\
	\kappa(\bm{r}) &= \sum_{k} V^{\dagger}_k(\bm{r})U_{k}(\bm{r}).\\
\end{aligned}
\end{equation}
 In DRHBc theory \cite{Zhou_PRC_2010}, the RHB equation \eqref{eqrhb} is solved by expanding quasiparticle spinors with the Dirac Woods-Saxon (DWS) basis \cite{Zhou_PRC_2003},
\begin{equation}
\varphi_{n\kappa m}(\bm{r}s)=\frac{1}{r}
\left(
\begin{array}{c}
	iG_{n\kappa}Y_{jm}^{l}(\hat{\bm{r}}s)\\
	-F_{n\kappa}Y_{jm}^{\tilde{l}}(\hat{\bm{r}}s)\\
\end{array}
\right),
\end{equation}
in which $n$ is the principal quantum number,
$\kappa=\pi(-1)^{j+1/2}(j+1/2)$ is a combination of the parity $\pi$ and the angular momentum $j$,
$m$ represents the projection of the angular momentum,
$G_{n\kappa}$ and $F_{n\kappa}$ are respectively the radial wavefunctions for large and small components of Dirac spinors,
and $Y_{jm}^{l}$ and $Y_{jm}^{\tilde{l}}$ are respectively the spin spherical harmonics with $l=j+\frac12\textrm{sgn}(\kappa)$ and $\tilde{l}=2j-l$. In order to describe the nucleus with axial deformation, the potentials and densities are expanded in terms of the Legendre polynomials \cite{Zhou_PRC_2003},
\begin{equation}
	f(\bm{r}) = \sum_{\lambda} f_{\lambda}(r) P_{\lambda}(\cos{\theta}),
\end{equation}
with
\begin{equation}
f_{\lambda}(r) = \frac{2\lambda+1}{4\pi}\int d\Omega f(\bm{r})P_{\lambda}(\cos{\theta}).
\end{equation}

\subsection{Finite amplitude method}

The random-phase approximation (RPA) equation is known to be equivalent to the time-dependent
Hartree-Fock (HF) equation in the small-amplitude limit \cite{Ring_Schuck}.
The Finite amplitude method is a practical method for solving the RPA equation in the self-consistent HF and density-functional theory \cite{Nakatsukasa_PRC_2007}.
The derivation and the implementation of FAM for the relativistic density functionals can be found in Ref. \cite{Liang_PRC_2013}.

For a nucleus slightly perturbed by an external field $\mathcal{F}(t)$ with the frequency $\omega$,
its generalized density $\mathcal{R}(t)$ and Hamiltonian $\mathcal{H}(t)$ will respectively oscillate around the equilibrium $\mathcal{R}_0$ and $\mathcal{H}_0$, in the small amplitude limit,
\begin{equation}
	\begin{aligned}
		&\mathcal{R}(t)=\mathcal{R}_0+\delta R(\omega)e^{-i\omega t}+\textrm{H.c.},\\
		&\mathcal{H}(t)=\mathcal{H}_0+\delta H(\omega)e^{-i\omega t}+\textrm{H.c.}.
	\end{aligned}
\end{equation}
In RPA, the induced density $\delta R(\omega)$ takes the contributions from creating (`20') and annihilating (`02') two quasiparticles  \cite{Ring_Schuck},
\begin{equation}
	\delta R(\omega)=\sum_{\mu\nu}\{X_{\mu\nu}(\omega)\beta^{\dagger}_{\mu}\beta^{\dagger}_{\nu}
	+Y_{\mu\nu}(\omega)\beta_{\nu}\beta_{\mu}\},
\end{equation}
in which $\beta^{\dagger}$ and $\beta$ are respectively the quasiparticle creating and annihilating operator,
and $X_{\mu\nu}(\omega)$ and $Y_{\mu\nu}(\omega)$ are the forward and the backward transition amplitudes relating to the quasiparticle pair $\mu\nu$.

Similarly, the induced Hamiltonian has the form,
\begin{equation}
	\delta H(\omega) =
	\frac12\sum_{\mu\nu}\{\delta H^{20}_{\mu\nu}(\omega)\beta^{\dagger}_{\mu}\beta^{\dagger}_{\nu}
	+\delta H^{02}_{\mu\nu}(\omega)\beta_{\nu}\beta_{\mu}\},
\end{equation}
where $\delta H_{\mu\nu}^{02}(\omega)$ and $\delta H_{\mu\nu}^{02}(\omega)$ are respectively the matrix element of the induced Hamiltonian.
As the term $\beta^{\dagger}\beta$ has no contribution at the RPA level, it is omitted here.

According to the equation of motion, $i\dot{\mathcal{R}}(t)=[\mathcal{H}(t)+\mathcal{F}(t),\mathcal{R}(t)]$, the following linear response equation can be obtained,
\begin{equation}\label{fameq}
\begin{aligned}
	&(E_{\mu}+E_{\nu}-\omega)X_{\mu\nu}(\omega)+\delta H_{\mu\nu}^{20}(\omega)= -F_{\mu\nu}^{20},\\
	&(E_{\mu}+E_{\nu}+\omega)Y_{\mu\nu}(\omega)+\delta H_{\mu\nu}^{02}(\omega)= -F_{\mu\nu}^{02}.\\
\end{aligned}
\end{equation}
Here the quasiparticle energy $E_{\mu}$ is the eigenvalue of  $\mathcal{H}_0$ and $F_{\mu\nu}^{20}$ and $F_{\mu\nu}^{02}$ denote the matrix element of the external field.

The induced Hamiltonian  $\delta H^{02}$ and $\delta H^{02}$  can be calculated from the variation of the single-particle Hamiltonian $\delta h$,
the variation of the paring field $\delta\Delta$ and $\delta\Delta^*$,
and the quasiparticle wavefunction $U$ and $V$ obtained in Eq. \eqref{eqrhb},
\begin{equation}\label{vh}
	\begin{aligned}
		&\delta H^{20}= U^{\dagger}\delta h V^* - V^{\dagger}\delta h^T U^*
		-V^{\dagger}\delta\Delta^*V^*+ U^{\dagger}\delta\Delta U^*,\\
		&\delta H^{02}= U^T \delta h^T V -V^T \delta h U
		-V^T \delta\Delta V +U^T \delta\Delta^* U.
	\end{aligned}
\end{equation}

The above equation is nothing but a representation transformation between a quasiparticle basis and a single particle basis.
Applying the same transformation to the induced density leads to,
\begin{equation}\label{vd}
\begin{aligned}
\delta\rho &= UXV^T+V^*YU^{\dagger},\\
\delta\kappa &= UXU^T+V^*YV^{\dagger},\\
\delta\kappa^* &=-VXV^T-U^*YV^{\dagger},\\
\end{aligned}
\end{equation}
where $\delta\rho$, $\delta\kappa$ and $\delta\kappa^*$ are respectively the variation of the single-particle density and the variation of the pairing tensor.
For the external field, the transformation reads
\begin{equation}
	\begin{aligned}
		&F^{20}=U^{\dagger}fV^*-V^{\dagger}fU^*,\\
		&F^{02}=U^{T}fV-V^{T}fU.\\
	\end{aligned}
\end{equation}

In FAM, the variation $\delta h$ ($\delta\Delta$ and $\delta\Delta^*$) are calculated from the single-particle Hamiltonian (the pairing field) at the perturbed density and the equilibrium $\rho_0$  ($\kappa_0$ and $\kappa_0^*$),
\begin{equation}\label{num_diff}
	\begin{aligned}
		\delta h&=\frac{1}{\eta}(h[\rho_0+\eta\delta\rho]-h[\rho_0]),\\
		\delta \Delta&=\frac{1}{\eta}(\Delta[\kappa_0+\eta\delta\kappa]-\Delta[\kappa_0]),\\
		\delta \Delta^*&=\frac{1}{\eta}(\Delta^*[\kappa_0^*+\eta\delta\kappa^*]-\Delta^*[\kappa_0^*]),\\
	\end{aligned}
\end{equation}
where $\eta$ is a small number used in the differentiation.

Starting with an initial guess $X^0_{\mu\nu}(\omega)$ and $Y^0_{\mu\nu}(\omega)$,
% Eq.  \eqref{vd}, \eqref{num_diff}, \eqref{vh}, and \eqref{fameq}
Eqs.  \eqref{fameq} to \eqref{num_diff} can be solved iteratively till convergence.
The converged amplitudes $X_{\mu\nu}(\omega)$ and $Y_{\mu\nu}(\omega)$ are used to get the strength function,
\begin{equation}
	S_F(\hat{F},\omega)
	=-\frac{1}{\pi}\mathrm{Im}\sum_{\mu\nu}\{F^{20*}_{\mu\nu}X_{\mu\nu}(\omega)
	+F^{02*}_{\mu\nu}Y_{\mu\nu}(\omega)\}.\\
\end{equation}

\section{Numerical details}

\begin{figure}	
	\centering
	\includegraphics[width=0.45\textwidth]{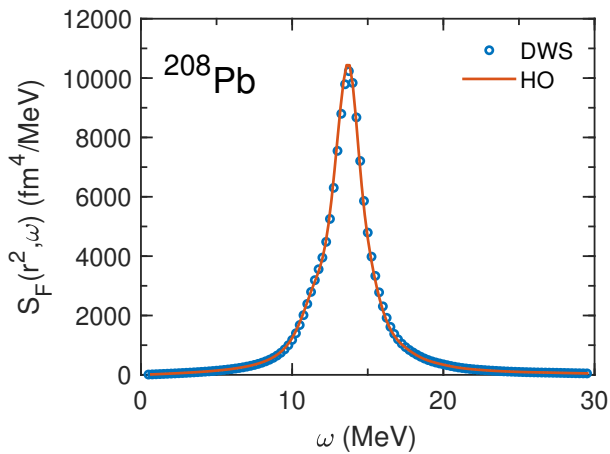}\\
	\caption{Strength functions of the ISGMRs for  $^{208}\textrm{Pb}$ calculated by expanding on DWS basis (circles) and HO basis (solid line).\label{num_chk}}
\end{figure}

The relativistic density functional PC-PK1 \cite{Zhao_PRC_2010} is used for the particle-hole channel,
which is well calibrated and gives accurate estimations of nuclei masses \cite{Zhao_PRC_2012,Lu_PRC_2015,Zhang_PRC_2021},
and shows excellent predicting power in a lot of nuclear phenomena like toroidal states \cite{Ren_NPA_2020},
magnetic rotations \cite{Zhao_PLB_2011,Wang_PRC_2017,Wang_PRC_2018},
antimagnetic rotations \cite{Zhao_PRL_2011},
multiple chirality in nuclear rotation \cite{Zhao_PLB_2017},
quadrupole moments \cite{Zhao_PRC_2014},
nuclear shape phase transitions \cite{Quan_PRC_2017},
and collision reactions \cite{Ren_PRC_2020}, etc.

In the DRHBc calculation, the numerical details suggested in Ref. \cite{Zhang_PRC_2020} are followed. A density-dependent zero-range pairing force with the strength $V_0=-325$ MeV is used for the particle-particle channel,
The relativistic Hartree-Bogoliubov equation is solved by expansion on a Dirac Woods-Saxon (DWS) basis \cite{Zhou_PRC_2003}. The DWS basis is constructed with a box size $R_{max} = 16$ fm, and the mesh size $\Delta r=0.1$ fm. The energy cutoff is $E_{cut} = 120$ MeV.

In the FAM calculations, the same numerical conditions are used.
With the efficiency of the FAM, a full two-quasiparticle (2qp) configuration space is constructed without any truncation.
For ISGMR, this means that the residual interactions among all the 2qp pairs with $K^{\pi}=0^+$ are considered.
To avoid possible singularity in Eq. \eqref{fameq}, a smearing width is added in the excitation energy, $\omega\rightarrow\omega+i\frac{\Gamma}{2}$.
If not mentioned otherwise, the smearing width is 2 MeV.
The parameter $\eta$ in Eq. \eqref{num_diff} to induce the numerical difference is set to $10^{-6}$.
The linear response FAM equation is solved iteratively.
The initial  amplitudes $X^0_{\mu\nu}$ and $Y^0_{\mu\nu}$ are set to be zero and the iteration is accelerated by the modified Broyden mixing method \cite{Johnson_RPB_1988}.
The numerical tolerance ($\max\{|\delta X/{X}|,|\delta Y/{Y}|\}$) for the iteration is $10^{-8}$.
The typical number of iterations varies from 20 to 50, depending on the excitation energy and the smearing width $\Gamma$ adopted.

In order to check the validity for the numerical implementation, the ISGMR strength function for $^{208}\textrm{Pb}$ is calculated and presented in Fig. \ref{num_chk}, in comparison with the result calculated in a harmonic oscillator (HO) basis with 20 shells by the code developed in \cite{Sun_PRC_2021}. Perfect agreements are achieved.
In both calculations, there are no truncation for the 2qp configuration space.

\section{ISGMR for the even-even calcium isotopes}

In the following, the DRHBc-FAM is applied to the even-even calcium isotopes $^{40-80}\textrm{Ca}$ to study effects of the continuum on the ISGMRs.

The $k$-th energy weighted moment relating to the monopole operator $r^2$ is defined as
\begin{equation}\label{mk}
	S_k=\int S_F(r^2,\omega)\omega^k d\omega.
\end{equation}
In particular, the energy weighted sum rule (EWSR) $S_1$ can be proved to be \cite{Lipparini_PR_1989},
\begin{equation}\label{EWSR}
	S_1^{0} = \frac{2\hbar^2}{m}A\langle r^2\rangle,
\end{equation}
with $A$ the mass number and $\langle r^2\rangle$ the mean-square radius.

From the energy weighted moment, the centroid energy,
\begin{equation}
%    E_c=\frac{S_1}{S_0}，
E_c=\frac{S_1}{S_0},
\end{equation}
which evaluates the position of a resonance peak, can be calculated.

In Table \ref{texp}, the centroid energies of ISGMRs for the even-even calcium isotopes $^{40-48}\textrm{Ca}$ are calculated and compared with the experimental data from Research Center for Nuclear Physics at Osaka University (RCNP) \cite{Howard_PLB_2020} and Cyclotron Institute at Texas A\&M University (TAMU) \cite{Youngblood_PRC_2001,Button_PRC_2017,Lui_PRC_2011}.
For the mass dependence of the centroid energies, the data from RCNP and TAMU show different trends.
Generally the calculated centroid energies are very close to the experimental data.
For $^{42,44,46}\textrm{Ca}$, the calculated centroid energies with DWS basis are slightly smaller than the calculations with HO basis. For $^{40}\textrm{Ca}$ and $^{48}\textrm{Ca}$, they are almost identical.
Since the centroid energy is related to the compression modulus \cite{Blaizot_PR_1980}, $E_c\sim\sqrt{K_A}$, an increasing $E_c$ with mass implies a positive value for the isospin asymmetry part of the incompressibility $K_{\tau}$, and vice versa. The trend of the calculated results agrees with the RCNP data, i.e., the centroid energy decreases with the mass number. Therefore, current calculations suggest a negative $K_{\tau}$, the same as the data from RCNP.

\begin{table}
	\caption{Calculated centroid energies (in MeV) of ISGMRs for $^{40-48}\textrm{Ca}$ by DWS basis and HO basis, in comparison with the experimental data of RCNP \cite{Howard_PLB_2020} and TAMU \cite{Youngblood_PRC_2001,Button_PRC_2017,Lui_PRC_2011}.\label{texp}}
	\begin{ruledtabular}
        \centering
		\begin{tabular}{lcccc}
			&\multicolumn{2}{c}{FAM calculations}&\multicolumn{2}{c}{experimental data}\\
			%\hline
			Nucl.&DWS&HO&RCNP&TAMU\\
			\hline
			$^{40}\textrm{Ca}$ &20.79  &20.80  &$20.2^{+0.1}_{-0.1}$ \cite{Howard_PLB_2020}&$19.2^{+0.40}_{-0.40}$ \cite{Youngblood_PRC_2001}\\
			$^{42}\textrm{Ca}$ &20.56  &20.61  &$19.7^{+0.1}_{-0.1}$ \cite{Howard_PLB_2020}&--\\
			$^{44}\textrm{Ca}$ &20.21  &20.31  &$19.5^{+0.1}_{-0.1}$ \cite{Howard_PLB_2020}&$19.50^{+0.35}_{-0.33}$ \cite{Button_PRC_2017}\\
			$^{46}\textrm{Ca}$ &19.86  &19.95  &-- &--\\
			$^{48}\textrm{Ca}$ &19.66  &19.66  &$19.5^{+0.1}_{-0.1}$ \cite{Howard_PLB_2020}&$19.9^{+0.2}_{-0.2}$ \cite{Lui_PRC_2011}\\
			\end{tabular}
	\end{ruledtabular}
\end{table}

\begin{figure}	
	\centering
	\includegraphics[width=0.45\textwidth]{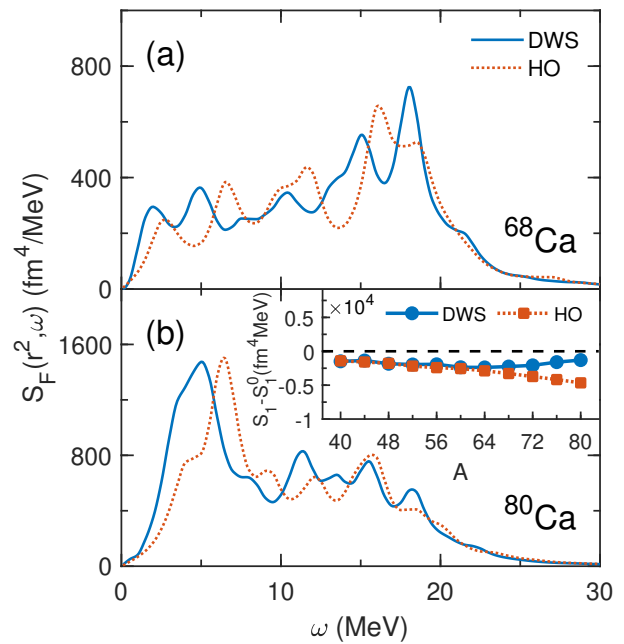}\\
	\caption{Monopole strength functions for $^{68}\textrm{Ca}$ (a) and $^{80}\textrm{Ca}$ (b) calculated with DWS basis (solid line) and HO basis (dotted line). In the inset, energy weighted sum rule for ISGMR is examined by the calculation up to 45 MeV with  DWS basis (circles) and HO basis (squares) for even-even calcium isotopes $^{40-80}\textrm{Ca}$.
\label{ewsr_cm}}
\end{figure}

Unlike the GMR for $^{208}\textrm{Pb}$ which concentrates in a single collective peak, the response functions of GMR for calcium isotopes are fragmented, thus are more dependent on the details of single-particle wave functions. As respectively illustrated in Fig. 2 (a) and Fig. 2 (b) for $^{68}\textrm{Ca}$ and $^{80}\textrm{Ca}$, the details of the response functions show differences between calculations with DWS basis and HO basis. Because, although the single-particle wavefunctions for the bound states are the same in both calculations, those for the continuum are different.
In the inset of Fig. \ref{ewsr_cm} (b), the energy weighted sum rule for ISGMR in Eq.(\ref{EWSR}) is examined in even-even calcium isotopes $^{40-80}\textrm{Ca}$.
The calculated results by DWS basis (circles) and by HO basis (squares) are presented.
The difference between the calculated $S_1$  and the model-independent $S_1^0$ are negligible.
For instance, for $^{72}\textrm{Ca}$, 98.2\% of the EWSR is exhausted below 45 MeV for DWS basis, and 96.8\% is for HO basis.
For the loosely-bound nuclei, calculations with DWS basis give slightly larger EWSR than that with HO basis because the coupling between the bound state and the continuum starts to work.
The spatial density distributions in exotic nuclei can hardly be described by HO basis unless extremely huge number of shells are used. In contrast, the DRHBc on the DWS basis with correct asymptotic behavior at the large distance from the center of the nucleus can achieve an equivalent performance as the calculations in the coordinate space \cite{Zhou_PRC_2003} for nuclear ground state properties.
Therefore, the calculations on DWS basis can take into account the continuum effects, and produces a value close to the  EWSR.
Another consequence of applying the HO basis to loosely bound nuclei is that, the spatial extension of the density at large radius is not well described, thus a too compact surface may be predicted. As a result, the energy of the soft monopole mode, which relates directly to the compression property of a nucleus near the surface, would be overestimated. For example, as presented in Fig. 2 (b) for $^{80}$Ca, the calculation with HO basis predicts a higher soft monopole mode than that with DWS basis.

\section{ISGMR for deformed and superfluid exotic nucleus $^{200}\textrm{Nd}$ }

To demonstrate the power of DRHBc-FAM, it is interesting to investigate the giant resonances in deformed and superfluid exotic nuclei.
We take the exotic nucleus $^{200}\textrm{Nd}$ with 60 protons and 140 neutrons as an example.
With the neutron chemical potential $\lambda_n=-0.94\textrm{~MeV}$ \cite{Zhang_PRC_2020}, the pairing correlation, deformation, and the continuum effect interplay in $^{200}\textrm{Nd}$,  and should be considered simultaneously.  In DRHBc calculations of such heavy deformed exotic nucleus, the box size of DWS basis is $R_{max} = 20$ fm, the energy cutoff is $E_{cut} = 150$ MeV, the angular momentum cutoff is $J_{cut}=23/2\hbar$ \cite{Zhang_PRC_2020}.

\begin{figure}
	\centering
	\includegraphics[width=0.45\textwidth]{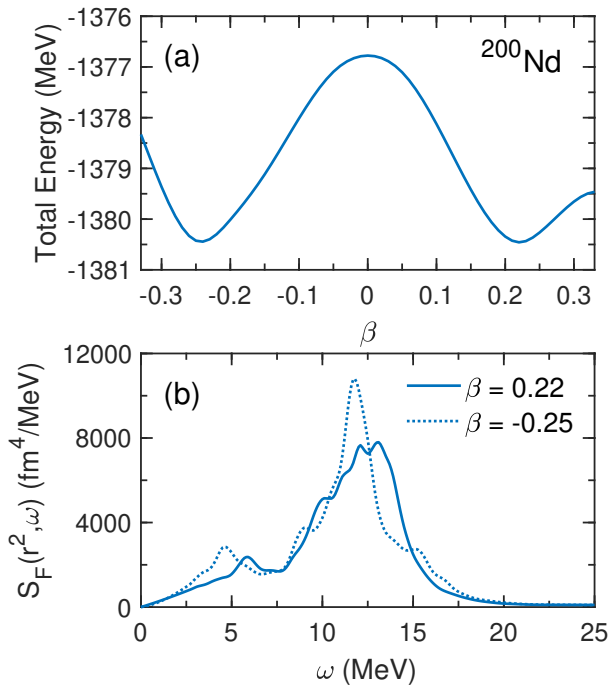}\\
	\caption{(a) Potential energy curve in $^{200}\textrm{Nd}$; (b) Strength function of the ISGMR built on the prolate isomer (solid line) and on the ground state (dotted line) in $^{200}\textrm{Nd}$ calculated by DRHBc-FAM.\label{nd200_gmr}}
\end{figure}

The  exotic nucleus $^{200}\textrm{Nd}$  locates at the prolate-oblate transition region in the neodymium isotopes with
$E = -1380.43\textrm{~MeV}$ at $\beta = 0.22$ and  $E = -1380.52\textrm{~MeV}$ at $\beta = -0.25$  in the potential energy curve, as shown in Fig. \ref{nd200_gmr} (a).

The giant monopole resonances for $^{200}\textrm{Nd}$ calculated by DRHBc-FAM are presented in Fig. \ref{nd200_gmr} (b). 
The strength functions are calculated up to 25 MeV with a step of 0.25 MeV, using a smearing parameter $\Gamma = 1$ MeV.
The main peaks of the ISGMR built on the prolate and on the oblate minima respectively locate around 12.5 MeV and 12.0 MeV.
Both ISGMRs are slightly broadened by the quadrupole deformations due to the well-known monopole-quadrupole coupling \cite{Yoshida_PRC_2010,Gupta_PRC_2016}.
The strength function shows a bump around 10 MeV at the low energy side of the main peak for the prolate case and a bump around 15 MeV at the high energy side for the oblate case, which turns out to coincide with the position of ISGQR ($K=0$) peak in the corresponding case.
For both prolate and oblate cases, soft monopole modes emerge at the low energy side of the strength function around 4.5$\sim$6 MeV.
For $^{200}\textrm{Nd}$, DRHBc-FAM calculations predict the soft monopole mode at 6.0 MeV for the prolate case, and at 4.5 MeV for the oblate case.
In the following, the structures of the soft monopole modes will be discussed.

\begin{figure}	
	\centering
	\includegraphics[width=0.45\textwidth]{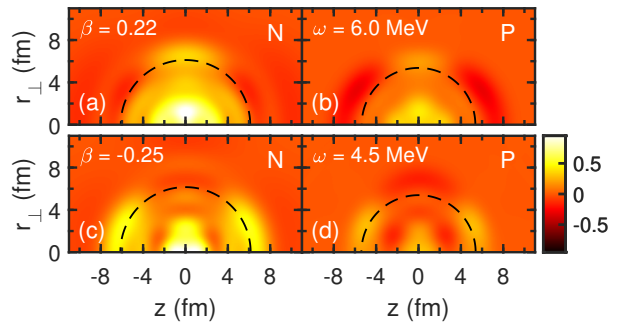}\\
	\caption{Transition densities of the soft monopole mode for $^{200}\textrm{Nd}$ in the prolate case for neutrons (a) and protons (b), and in the oblate case for neutrons (c) and protons (d). The dashed line in each panel indicates the root-mean-square radius of neutron or proton density.\label{nd200_td}}
\end{figure}

A straightforward reflection of the nucleus vibration is the transition density defined as,
\begin{equation}
	\delta\tilde\rho(\omega,r_{\perp},z)
	= \eta\textrm{Im}\sum_m\sum_{n\kappa,n'\kappa'}
	   \varphi_{n\kappa m}^{\dagger}(\bm{r})
	   \delta\rho_{n\kappa,n'\kappa'}^m(\omega)
	   \varphi_{n'\kappa' m}(\bm{r}),
\end{equation}
in which $\delta\rho_{n\kappa,n'\kappa'}^m(\omega)$ denotes the matrix element of the induced single-particle density in the DWS basis.
In Fig. \ref{nd200_td}, the normalized transition densities of the soft monopole mode for $^{200}\textrm{Nd}$ are illustrated for neutrons (a) and protons (b) at 6.0 MeV in the prolate case, as well as for neutrons (c) and protons (d) at 4.5 MeV in the oblate case.
Because of the deformation, the transition densities are anisotropy in the intrinsic frame of reference.
The nucleus vibrates differently in the $z$-direction and in the $r_{\perp}$-direction.
The root-mean-square radii of $^{200}\textrm{Nd}$ are respectively 5.90 fm for the prolate case, and 5.93 fm for the oblate case.
Near the surface, the nucleons may vibrate in a different phase with respect to the nucleons in the core.
For neutrons in the prolate case in  Fig. \ref{nd200_td}(a), the out-of-phase vibrations can be identified in the $z$-direction.
For neutrons in the oblate case in  Fig. \ref{nd200_td}(c), it occurs in the $r_{\perp}$-direction.
The situations for proton transition density are similar to their corresponding neutron cases but with smaller amplitudes.

In axial deformed case, the total angular momentum $J$ is no longer a good quantum number.
The mixing between the monopole vibration with $J=0$ and the quadrupole vibration with $J=2$, or even higher order multipole vibrations may occur.
To investigate the structure of the soft monopole mode, the contribution from different $J$ components to the transition density are analyzed in the following.
The angular momentum projection of the intrinsic transition density can be performed as \cite{Niksic_PRC_2013},
\begin{equation}
    \delta\rho^{J}(\omega,\bm{r}) = \delta\rho^{J}(\omega,r)Y_{JK}(\Omega),
\end{equation}
where the radial projected transition density is defined as,
\begin{equation}
\delta\rho^{J}(\omega,r)=\int d\Omega\delta\tilde\rho(\omega,r_{\perp},z)Y_{JK}(\Omega).
\end{equation}
For ISGMR, the $z$-component of the angular momentum $K=0$.

\begin{figure}	
	\centering
	\includegraphics[width=0.45\textwidth]{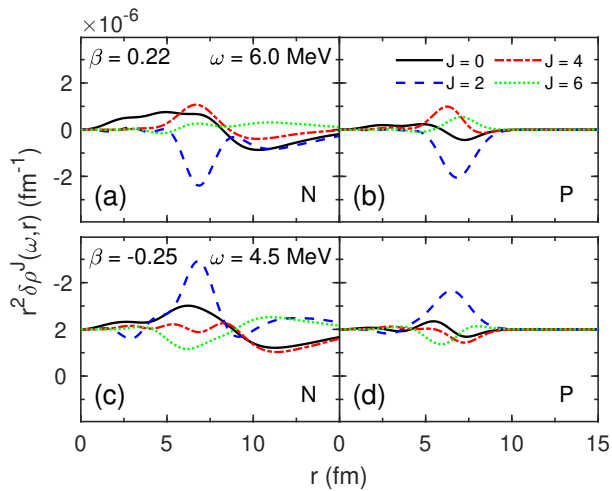}\\
	\caption{Radial distributions of the projected transition densities for the soft monopole modes in $^{200}\textrm{Nd}$,
		in the prolate case for neutrons (a) and protons (b), and in the oblate case for neutrons (c) and protons (d).
		\label{nd200_pj}}
\end{figure}

In Fig. \ref{nd200_pj}, the radial distributions of the transition densities in the prolate and oblate cases for neutrons and protons are presented.
In general, the contributions from $J=0$ and $J=2$  dominate in the transition density, especially in the core region.
The components with $J=4$ and $J=6$ make minor contributions and are not negligible.
For example, in the oblate case, the contribution of $J=4$ part counteracts with that of the $J=2$ part at $r\approx 3$ fm.
The vibration of neutrons surpasses that of protons in the inner part of the nucleus, and extends to larger distance.
To be specific, the vibrations of neutrons in  Fig. \ref{nd200_pj}(a) and (c) extend as far as 15 fm, while those of protons in  Fig. \ref{nd200_pj}(b) and (d) decay quickly around 8 fm.
The long-tail of the neutron transition density manifests the loosely-bound nature of $^{200}\textrm{Nd}$.
Thanks to the DRHBc which describes the asymptotic behavior of the wave functions at large $r$ and treats the continuum more accurately, the long-tail of the neutron transition density is well described.

Comparing the soft monopole modes built on the prolate shape isomer and on the oblate ground state, obvious distinctions exist between their behaviors in the surface region at $r\approx6$ fm.
In the prolate case, the $J=2$ and the $J=0$  neutron transition densities are out-of-phase.
In the oblate case, the $J=2$ and the $J=0$  neutron transition densities are in-phase.
Although the $J=4$ and $J=6$ parts counteract with $J=2$ part, but they are much smaller in amplitude.
Therefore, the quadrupole part with $J=2$ dominates the vibrations near the surface, and generates the in-phase or  out-of-phase vibrations for the neutrons near the surface.

\section{Conclusion}
In this work, finite amplitude method is implemented on deformed relativistic Hartree-Bogoliubov theory in continuum.
The DRHBc-FAM is validated by comparing the calculated ISGMR for $^{208}\textrm{Pb}$ with the result by the existing code on HO basis.
The ISGMRs for even-even calcium isotopes are calculated, and a good agreement with the experimental centroid energies is obtained.
For the loosely bound calcium isotopes like $^{68}\textrm{Ca}$ and $^{80}\textrm{Ca}$, the DRHBc-FAM calculated results are closer to the EWSR than the calculations on HO basis.

As both the continuum effect and deformation are considered simultaneously in DRHBc-FAM, an illustrative example is presented for the deformed exotic nucleus $^{200}\textrm{Nd}$.
For $^{200}\textrm{Nd}$, the prolate shape and the oblate shape coexist and a soft monopole mode near 6.0 MeV is found in the prolate case, and another one near 4.5 MeV is found in the oblate case.
For the soft monopole mode of $^{200}\textrm{Nd}$, the vibration of neutrons is much stronger than that of protons.
Since $^{200}\textrm{Nd}$ is loosely bound, the neutron transition density extends to very far.
Near the surface region, $J = 0$ part and $J = 2$ part neutron transition densities are destructive in the prolate case, and are constructive in the oblate case.

\begin{acknowledgments}
We thank C. Pan, K. Zhang, D. Vretenar, and T. Nik\ifmmode \check{s}\else \v{s}\fi{}i\ifmmode \acute{c}\else \'{c}\fi{} for helpful discussions. This work is partly supported by the National Key Research and Development Program of China (Grants No. 2018YFA0404400 and No.2017YFE0116700), the National Natural Science Foundation of China (Grants No. 11621131001, No. 11875075, No.11935003, and No. 11975031), the State Key Laboratory of Nuclear Physics and Technology, Peking University (Grant No. NPT2020ZZ01), and the China Postdoctoral Science Foundation under Grant No. 2020M680182. This work is supported by High performance Computing Platform of Peking University.
\end{acknowledgments}

% ---- bibliography ----

%\bibliography{author_Journal_year}

\end{document}